# Forced Changes Only: A New Take on the Law of Inertia

DANIEL.HOEK@VT.EDU, VIRGINIA TECH, DANIELHOEK.COM, NOVEMBER 2021[1]

ABSTRACT: Newton's First Law of Motion is typically understood to govern only the motion of force-free bodies. This paper argues on textual and conceptual grounds that it is in fact a stronger, more general principle. The First Law limits the extent to which *any* body can change its state of motion — even if that body is subject to impressed forces. The misunderstanding can be traced back to an error in the first English translation of Newton's *Principia*, which was published a few years after Newton's death.

On the received view of Newton's First Law of Motion, the law is exclusively concerned with the motion of *force-free* bodies, stating that such bodies continue their state of rest or uniform motion in a straight line. For instance, distinguished Newton scholar Brian Ellis formulates the law as follows: "every body not subject to the action of forces continues in its state of rest or uniform motion in a straight line" (Ellis 1965, 35). Richard Feynman's lectures state it thus: "if an object is left alone, is not disturbed, it continues to move with a constant velocity in a straight line if it was originally moving, or it continues to stand still if it was just standing still" (Feynman et al. 1963, §6.2). Or take Thomas Kuhn: "In the absence of an external force applied to it, a body moves continuously at constant speed in a straight line" (2000, 68).

The thesis of this paper is that such paraphrases of Newton's First Law are all incorrect, because the law, as Newton stated it, is not just a description of the motion of force-free bodies. It is in fact a stronger, more general principle, constraining the motion of all bodies. Here is Newton's own formulation of the law, in the authoritative Cohen and Whitman translation (Newton 1999):

> Every body perseveres in its state of being at rest or of moving uniformly straight forward, except insofar as it is compelled to change its state by the forces impressed.    ($L_I$)

As I understand it, the phrase "except insofar" (*nisi quatenus*) does not exempt bodies that are subject to impressed forces from the scope of the law altogether. Instead it puts a limit on the extent to which the state of motion of any body can change: it changes *only insofar as the impressed forces compel it to*. This principle rules out changes in speed or direction due to influences besides impressed forces, like loss of impetus, "natural motion" or spontaneous animation in living organisms. It also bars changes in motion that are not compelled by anything at all, like the random swerve of Lucretian atoms. Those restrictions apply to *all* massive bodies, whether they are force-free or subject to impressed forces. So a better way to paraphrase the First Law would be this: "Every change in a body's state of motion is due to impressed forces," or alternatively: "Bodies only accelerate by force."

When presenting or discussing this reading of the First Law $L_I$, I typically get one of two reactions. Some find my reading too wild and unconventional to take seriously. Others think that it is so obviously correct that it is barely worth arguing for. Obvious or not, the strong reading of the First

---

[1] Forthcoming in *Philosophy of Science*. Thanks to Zvi Biener, Eliyah Cohen, Nico Kirk-Giannini, Jordan MacKenzie, Tim Maudlin, Wendy Parker, Lydia Patton, Gideon Rosen and Trevor Teitel for their insightful comments on earlier drafts. I am also very indebted to two anonymous referees for *Philosophy of Science*, whose scholarly insight and generous enthusiasm led to significant improvements.

Law has, to the best of my knowledge, never been clearly articulated or explicitly defended in print. By contrast, weak readings of the law have been explicitly endorsed by scores of prominent historians, philosophers, and physicists, as is extensively documented below. So I think it is worthwhile to set the record straight by articulating the strong reading in this explicit manner, and by laying out the evidence in its favour. That will be the task of this paper.

Section I adduces the main textual evidence for the strong reading, arguing that the presentation of the First Law in the *Principia* strongly favours it. In particular, Newton elucidates the law with some examples, including one about a spinning top. Since the parts of this top are neither force-free nor move in straight lines, this illustration only makes sense on a strong reading of the law. In Section II, I further argue that a weak First Law would leave the formulation of Newtonian Mechanics woefully incomplete. Even when the Second Law is considered, we still need a strong First Law to rule out unforced fluctuations in bodies' states of motion. Section III shows how this resolves a long-standing exegetical problem I call the *Independence Problem*. This is the difficulty of explaining why Newton postulated a separate First Law at all, given that the Second Law appears to entail it.

The second half of the paper (which a hasty reader may skip) places these findings in context. Section IV concerns the origin of the traditional, weak reading of the First Law, and hypothesises that it can be traced back to a faulty English translation by Andrew Motte from 1729. Section V examines the pervasive impact that this reading has had on the academic literature about inertia. To do so, we turn to another puzzle, the *Triviality Problem*. This problem turns on the observation that, on a weak reading, the First Law is a generalisation with no instances, since there are no force-free bodies in nature (if only because of gravitation). The issue does not arise on the strong reading of the law. After all, the principle that every change in bodily motion is compelled by a force has many instances: it is instantiated every time a body changes its speed or direction. Thus the considerable impact of the Triviality Problem is a testament to the influence of the weak reading of the First Law.

## I. The Text

Recall once more Newton's formulation of his First Law:

> Every body perseveres in its state of being at rest or of moving uniformly straight forward, except insofar as it is compelled to change its state by the forces impressed. [2]  ($L_I$)

The question before us is whether Newton intended the traditional, weak reading of this formula, $W_I$, or the strong reading $S_I$ that I am advocating:

> If no forces are impressed on a body, it will persevere in its state of uniform rectilinear motion.  ($W_I$)

> Every change in a body's state of uniform rectilinear motion is compelled by the forces impressed on that body.  ($S_I$)

---

[2] The Latin reads: *Corpus omne perseverare in statu suo quiescendi vel movendi uniformiter in directum, nisi quatenus illud a viribus impressis cogitur statum suum mutare.* Newton added the word "*illud*," ("it") for emphasis in the third edition; otherwise the formulation of the First Law is the same in all three editions of the text. For a comprehensive account of the development of this formulation in Newton's prior work, and its precursors in the work of Galileo, Descartes and Gassendi, see Herivel 1965, §1.4 and Ch. 2. Except where indicated otherwise, all *Principia* quotations in this paper are from Newton 1999 (English) and Newton 1726 (Latin).

To put it in terms of momentum, $W_I$ says a body maintains its *momentum* if no forces are impressed on it, while $S_I$ says that every change in a body's *momentum* is due to impressed forces. $S_I$ entails $W_I$, since it entails that a body on which no forces are impressed does not change its state of motion. But $W_I$ and $S_I$ are not contrapositives: $W_I$ fails to entail $S_I$, because $W_I$ leaves open the possibility of an unforced change in the state of motion of a body subject to impressed forces (it could be subject to influences other than impressed forces, or undergo random changes, for no reason at all).

At a glance, $W_I$ may seem like an acceptable paraphrase of $L_I$. But as discussed in the introduction, that impression dissipates once we ask what $L_I$ tells us about bodies that are subject to an impressed force. It is intuitively clear that it does say something: namely that their state only changes *insofar* as the forces impressed compel that state to change (and no more). The difference between $S_I$ and $W_I$ is precisely that $S_I$ captures this limitation on the extent to which bodies can change their state of motion, while $W_I$ ignores it. This consideration alone establishes $S_I$ as a credible alternative to $W_I$.

Newton uses similar wording in setting the stage for the First Law in the *Principia*'s definitions section. A different way of saying that a body *diverges* from its state of motion only to the extent that it is compelled, is to say that a body *maintains* its state of motion *to the extent that it can*:

> [E]very body, *so far as it is able*, perseveres in its state either of resting or of moving uniformly straight forward.[3]

Newton took the crucial proviso "so far as it is able" (*quantum in se est*) from Descartes. As this phrase implies, not all bodies do persevere in their state of motion. But again, that clearly does not limit the scope of the statement to unperturbed bodies. *All* bodies persevere insofar as they are able: it is just that some are more able than others.[4] The First Law $L_I$ adds to this claim by saying what curbs bodies' ability to persevere in their uniform rectilinear motions: namely the action of impressed forces.

Immediately following the statement $L_I$ of the First Law, Newton gives three concrete examples that are intended to elucidate how this law can be applied in practice. First, he points out that "Projectiles persevere in their motions, except insofar as they are retarded by the resistance of the air and are impelled downward by the force of gravity." The claim that the projectile would move uniformly in a straight line, but for the presence of impressed forces, is warranted on both the weak and strong readings of the First Law. Newton's second example is more probative for our purposes:

> A spinning hoop [top], whose parts by their cohesion continually draw one another back from rectilinear motions, does not cease to rotate, except insofar as it is retarded by the air.

Here, Newton claims that without air resistance, the parts of a spinning top would keep on going in circles, because they are continually drawn to the centre. But clearly, they would still diverge from uniform rectilinear motion — so how is this an application of the First Law $L_I$? Well, the parts of the spinning top do diverge from uniform rectilinear motion, but *only insofar* as they are compelled to do so by the forces impressed on them. With this example, Newton shows that the First Law also applies to bodies that are subject to impressed forces. But this would not be so if the law were the weak principle $W_I$, which says nothing at all about the motion of bodies subject to impressed forces. This clearly indicates Newton must have had the strong principle $S_I$ in mind. Indeed, it is not a stretch to speculate that his purpose in adducing this particular example was to clarify that very point.

---

[3] My emphasis. This passage is from Definition 3, the characterisation of inherent or inertial force (*vis insita* or *vis inertiae*). For discussions of Newton's concept of inherent force, and of its connection to the First Law, see McGuire 1994, Cohen 1999, §4.7-8, Garber 2012.

[4] As a reviewer points out, an analogous point applies to the *prima lex naturae* of Descartes' *Principia Philosophiae*: "Any particular thing, *so far as it is able*, always perseveres in the same state" (Part 2, §37). This generalisation, too, quantifies over all things — not just unperturbed objects. What Newton's statement $L_I$ chiefly adds to the Cartesian understanding of inertia is a positive conception of what *does* perturb bodies' states of motion: namely impressed forces. For more on the phrase "*quantum in se est*" and its Cartesian roots, see Cohen 1964.

This observation is further reinforced by Newton's third and final illustration:

> And larger bodies — planets and comets — preserve for a longer time both their progressive and their circular motions, which take place in spaces having less resistance.

Here, Newton employs the First Law to explain a contrast: the planets and comets preserve their elliptical orbital motion over very long periods of time, unlike projectiles and spinning tops on earth. Newton's explanans is that these bodies, travelling through empty space, meet with fewer resistance forces. This is an application of $S_I$, which entails that any divergence of the planets from their regular orbits would have to be due to further forces. But again, no such conclusion can be derived from $W_I$, which concerns only force-free bodies on straight trajectories.

The two passages just cited carry great exegetical weight, because Newton puts these examples forward precisely in order to elucidate the content and application of the First Law. Thus they count very strongly in favour of the strong reading $S_I$, and very strongly against the traditional weak interpretation $W_I$. The latter has the highly uncharitable consequence that Newton manages to misapply his own First Law twice *within his own elucidation of that very law*.

This last point was clearly recognised by E.J. Dijksterhuis (1963, 466-7; 1996, 513) — but astonishingly Dijksterhuis *does* draw the uncharitable conclusion. Michael Wolff is also tempted:

> What does a rotating spinning top even have to do with clarifying the First Law? What does its rotational motion … have to do with the inertial condition of the First Law? A uniform linear motion is no more performed by the rotating top than it is properly at rest. All parts of its body undergo accelerations in the relevant sense … the spinning top example appears to be nothing more than an unfortunate error of judgement by Newton. Dijksterhuis thinks it simply does not belong here. (Wolff 1978, 324-5, transl. from German)

Wolff, however, ultimately stops short of fully endorsing Dijksterhuis' view, opting instead for the questionable suggestion that the motion of the parts of a spinning top is, in a sense, rectilinear "within each moment" (Ibid., 328). It is a testament to the profound entrenchment of the weak reading of the First Law that these careful scholars should sooner question Newton's understanding of his own law, or describe the parts of a spinning top as being in rectilinear motion, than adjust or even reconsider their own reading of that law (cf. also Cohen 1999, Ch. 5; Ludwig 1992).

This striking loyalty to the weak reading $W_I$ should be viewed and understood in its proper context, as part of a powerful and firmly established tradition that goes back almost three centuries (as we shall see in Section IV). At the start of the paper, I cited Richard Feynman, Brian Ellis, Thomas Kuhn and Ernst Mach, all unambiguously embracing traditional, weak readings of the law. It takes little effort to find similar quotes from other notable scholars, philosophers and physicists, even if we restrict attention to articles specifically about the content of the First Law.[5] One such article, by Herbert Pfister, perfectly captures the general consensus: "*Any* formulation of Newton's first law is in essence of the type that some special class of physical objects (so called free particles) move on a special class of mathematical curves (so called straight lines)" (Pfister 2004, 51, my emphasis).

But Newton's own formulation of the First Law is not of this type, if the foregoing is correct. Firstly, the scope of the law is not restricted to any special class of bodies. Secondly, the law concerns the extent to which bodies *diverge* from motion along straight lines. I have so far argued that this strong reading $S_I$ of the First Law is forced on us by Newton's own formulation and elucidation of his First Law. Let me now put forward a further argument in its favour.

## II. The Second Law

One way to compare the weak and strong readings $W_I$ and $S_I$ is to ask the straightforwardly exegetical question: which of these principles did Isaac Newton actually have in mind when he wrote his First Law? In my view, the textual evidence adduced in the last section settles that question in favour of the strong reading $S_I$. But we can also ask a different question: which reading yields a better formulation of Newtonian mechanics? As I shall argue, $S_I$ compares favourably to $W_I$ in this respect too. To see this, we will need to discuss the role of the Second Law of Motion.

While the First Law says changes in a body's state of motion are due to forces, the Second describes just how a body's state of motion is changed by a given force:

> A change in motion is proportional to the motive force impressed and takes place along the straight line in which that force is impressed. [6]             ($L_{II}$)

In the first clause, Newton uses the term *motive force* rather than *impressed force*, because he is referring to the scalar quantity that measures the strength of the impressed force; the impressed force has not only a strength but also a spatial direction.

To the modern eye, it is tempting to read $L_{II}$ as stating that the body's overall change in motion is proportional to the total or net force impressed (more on this reading in Section III). But that is not what Newton meant. He speaks of a singular "force" in $L_{II}$, contrasted with the plural "forces" of $L_I$. This shift indicates that $L_{II}$ describes effect of a single, individual force by itself, rather than the overall effect of impressed forces. Accordingly, the "change in motion" that $L_{II}$ describes is a "change in motion *due to a particular impressed force*" (such as friction or gravity). As Newton puts it in his explanation of the Second Law, the law describes the "new motion" contributed by a force. In modern terms, $L_{II}$ describes the vector component of a body's acceleration associated with a particular force, and not the overall acceleration.

---

[5] An article on the First Law by historian of science Norwood Russell Hanson could not be more emphatic: "Newton's first law of motion reads: EVERY BODY FREE OF IMPRESSED FORCES EITHER PERSEVERES IN A STATE OF REST OR IN UNIFORM RECTILINEAR MOTION *AD INFINITUM*" (Hanson 1965, 7, his emphasis). The influential treatment of inertia by philosophers John Earman and Michael Friedman puts it in more technical terms: "the space-time trajectories of particles free of impressed forces are affine geodesics" (Earman and Friedman 1973, 334). Or take a widely cited piece by physicist James L. Anderson: "The first law describes the behaviour of bodies not acted on by external forces" (Anderson 1990, 7). Besides Ellis, Hanson, Wolff and Dijksterhuis, historians that have endorsed weak readings of the First Law include Alexandre Koyré (1950, 260-1), Arnold Koslow (1969, 551-2), and I. Bernard Cohen (1980, 106-9). For other philosophers, see Russell 1903, Ch. 56; Stein 2002, 275; Brown 2006, §2.2. For more physicists, see Hertz 1899, 178; Eddington 1929, 124; Rigden 1987; Barbour 1989, 5, 20-1. For a survey of physics textbooks, see Galili and Tseitlin 2003. See also Section V.

[6] *Mutationem motus proportionalem esse vi motrici impressæ, & fieri secundum lineam rectam qua vis illa imprimitur.*

In the exposition of *Principia*, the entire concept of a total or net force is in fact several steps downstream from Second Law $L_{II}$. Newton only introduces it in Corollary 2 of the laws, which states that we may treat a multiplicity of forces as composing a singular "direct force," and conversely that a single force can be decomposed ("resolved") into multiple component forces. This result is based on Corollary 1, which describes the effect of two forces acting simultaneously:

> A body acted on by [two] forces acting jointly describes the diagonal of a parallelogram[7] in the same time in which it would describe the sides if the forces were acting separately.

To prove this claim, Newton invokes the Second Law $L_{II}$ to describe the component changes produced by each of the two impressed forces individually, and shows how these two changes combine to produce the described effect — which is an overall change in motion that is neither proportional nor parallel to either of the two forces acting. He also cites the First Law $L_I$ to show that there is no further displacement beyond the change compelled by the two impressed forces.[8]

If Newton's Second Law $L_{II}$ were itself already a description of the effect of multiple forces acting, there would be no need to treat the case of two forces in a separate corollary — much less to cite the First Law in doing so. The derivation of Corollary 1 only makes sense if we take it that the "change in motion" described by $L_{II}$ is compatible with the existence of further simultaneous changes in motion. That is to say, $L_{II}$ describes how a particular impressed force diverts a body from its uniform rectilinear trajectory, in a way that is meant to be consistent with the possibility that other impressed forces should simultaneously divert it further.

But — and this is the crucial part — it is the First Law, not the Second, which tells us that any additional diversions from the state of motion would have to be due to further impressed forces. $L_{II}$ just describes the change in motion produced by an impressed force. Only the First Law rules out the possibility that something other than an impressed force should produce a further change in motion. Moreover, when impressed forces are acting, as in the situation described in Corollary 1, we need the strong reading $S_I$ of the First Law in order to rule out that possibility.

Pre-Newtonian theories of mechanics posited all sorts of reasons besides forces for changes in speed or direction. According to Buridan, projectiles shot into the air start to fall when they run out of *impetus*. Aristotle said that heavenly bodies, made of aether, naturally move in circles, while bodies made of earth naturally bend towards the centre of the universe. Animate bodies were thought to move spontaneously, of their own volition. And Lucretius thought atoms could pivot and swerve entirely at random, for no reason at all. Plainly, Newton's vision of mechanics was inconsistent with all these potential ways for a body's state of motion to change.[9] But the Second Law $L_{II}$ does not rule them out, as we just saw. Clearly the Third Law does not do so either. So this job falls entirely to the First Law.

---

[7] From a contemporary perspective, Newton's claim that the body travels along the *diagonal*, a straight line, seems incorrect. However, as Brian Ellis (1962) pointed out, strictly speaking, both $L_{II}$ and Corollary 1 concern instantaneously applied forces or *impulses* — "kicks". Continuously acting forces are mostly described as rapid sequences of small kicks (but cf. Pourciau 2006). For present purposes, this difference between Newton's statement of the Second Law and modern, 'rate of change'-based formulations does not matter.

[8] Ernst Mach criticised this derivation of Corollary 1 on the grounds that it smuggles in the assumption that the two forces act independently (Mach 1919, 197). But this is not an additional assumption: it follows from $L_{II}$, which says that the effect of a force is *always* proportional to its strength (whether or not another force is acting).

It follows that a weak reading $W_I$ of the First Law would leave Newtonian mechanics incomplete. Even in combination with the other laws, $W_I$ is too weak to derive the Newtonian thesis that the forces acting on a body determine the evolution of its state of motion. Both $W_I$ and $L_{II}$ leave open the possibility of unforced changes in bodies' states of motion — except in the special case of force-free bodies. That is hardly a consolation, given that there are no truly force-free bodies (see Section V). Only with the strong First Law $S_I$ do we get a formulation of Newtonian Mechanics that rules out unforced changes in the states of motion of all massive bodies.

That is not to say that $W_I$ is altogether inert. It is strong enough to contradict ancient and medieval theories of mechanics. According to Aristotle, the moon completes its natural circular motions without any forces acting on it; and according to Buridan's impetus theory, projectiles fall down without the aid of any forces. In his discussion of Newton, Tim Maudlin (2001, Ch. 1) locates the core significance of the First Law in this discontinuity with the past (cf. also Koyré 1965, Cohen 2002). According to Maudlin, the import of the First Law consists in the radically egalitarian thought that *every* body, on earth and in the heavens, has the same natural motion: namely uniform motion in a straight line.

I am sympathetic to this idea, but I think it slightly understates how radical the First Law of motion is. Newton did not just deny the possibility of acceleration in the absence of impressed forces, as $W_I$ does. For that would leave the door ajar to impetuses, natural motions, spontaneous actions, random fluctuations and the like: they may yet rear their heads in the *presence* of impressed forces. In fact, Newton's First Law banished those influences altogether by insisting that *only* impressed forces can change a body's state of motion.

## III. The Independence Problem

It is frequently noted that Newton's "Laws or Axioms" of Motion appear to contain a redundancy. In the words of W.W. Rouse Ball, the First Law "seems to be a consequence of the second law, and if so it is not clear why it was enunciated as a separate law" (1879, 77). L.W. Taylor is more assertive: "the first law could have been stated more logically as a corollary to the second law, which it is, rather than as an independent law which it most certainly is not" (1959, 131). And Thomas Kuhn writes: "Newton's first law is a consequence of his second, and Newton's reason for stating them separately has long been a puzzle" (2000, 68). Let us call that puzzle the *Independence Problem*.

Rouse Ball and Taylor offer no solution to the puzzle. Kuhn suggests that Newton's reasons must have been pedagogical in nature: "Separating [the laws] to the extent possible displayed the nature of the required changes more clearly" (ibid.). I. Bernard Cohen claims that the problem can be solved by saying that the First and Second Law concern different kinds of impressed force, but does not develop that suggestion in much detail (Cohen 1999, 110).

---

[9] Newton did think our thoughts could move our bodies — but only through the mediation of an impressed force, perhaps related to electricity or gravity (see Dempsey 2006).

The discussion above yields a less contrived and altogether more satisfying response to the Independence Problem. For on the present reading of Newton's First and Second Laws, it is simply not the case that the Second Law entails the First. The first two laws are, roughly, converses of one another, and so they are logically independent. The Second Law says that every impressed force compels a proportional change in motion in the direction of the force, while the First says that every change in motion is compelled by a force. The former fails to entail the latter, and so the Independence Problem vanishes once Newton's laws are understood in this way.[10]

Additional confirmation for this way of understanding the separate roles of the First and Second Law can be found in the way Newton puts his Laws of Motion to use in the *Principia*. When calculating the diversion from rectilinear motion produced by a given force, he cites the Second Law. But when starting that a given diversion requires an impressed force, he always cites the First Law, not the Second. Thus the First Law sustains the inference from acceleration to force, while the Second takes us from force to acceleration.[11]

Why can't the Second Law $L_{II}$ do the work that Newton allots to the First? $L_{II}$ says that a change in motion is proportional to the force impressed. Doesn't it follow from this that the presence of a change in motion implies the presence of an impressed force? I claim it does not. Note that by parallel reasoning, the *absence* of acceleration would imply the *absence* of impressed forces. However, that is not so: when equal and opposite forces act on a body simultaneously, the resulting accelerations cancel each other out. Acceleration can be absent even in the presence of forces, consistently with the Second Law $L_{II}$. In much the same way, $L_{II}$ is by itself consistent with the possibility of acceleration in the absence of forces, because it is consistent with unforced accelerations of various kinds — this is what I argued in Section II. It is the First Law $L_{I}$, and only the First, which rules those out.

It is worth noting that this applies to the Second Law $L_{II}$ as Newton formulated it. It does not carry over to the Eulerian formulation $\mathbf{F} = m \cdot \mathbf{a}$ that most modern physics texts employ (e.g. Feynman et al. 1963, §9.2; Resnick et al. 2002; Thorne and Blandford 2017):

> $\mathbf{F} = m \cdot \mathbf{a}$, where $\mathbf{F}$ denotes the vector sum of all the forces acting on a body (the *net force*), and $\mathbf{a}$ denotes the overall acceleration of that body.   (FMA)

Equivalently, FMA can be formulated in a slightly more Newtonian way as $\mathbf{F} = \frac{d\mathbf{p}}{dt}$, where the right hand side represents the overall momentum change of the body. Either way, there remains a crucial difference between Newton's $L_{II}$ as I understand it and FMA. The former relates *individual* impressed forces to *components* of the body's acceleration while latter relates the *net* force acting on a body to its *overall* acceleration.

---

[10] A referee outlined another plausible response to the Independence Problem. More than one response could be correct: there may be multiple ways in which the First and Second Laws complement one another. I just claim that my reading of Newton gives you one solution — it need not be the only one.

[11] Newton directly cites the Second Law in the following *Principia* proofs: corr. 1, 3, 5 and 6 of the laws, prop. 6 of book 1 (1st ed.) and props. 3, 8 and 24 of book 2 (Newton 1999, 64, 66, 69, 100*n*, 281, 287, 296, 346). In three instances (corr. 3, 5 and 6), this is to show that equal forces produce equal changes in motion. In all other instances, it is to establish the direction and/or magnitude of the acceleration due to a given force. Newton cites the First Law in the proofs of corr. 1, of prop. 1 in book 1, prop. 32 of book 2, and prop. 17 of book 3 (ibid., 64, 90, 370, 466). In all cases, this is to establish that an impressed force would be necessary to produce an acceleration.

Once this difference in content is recognised, we can see that FMA has various entailments that $L_{II}$ lacks. It captures what, from an empirical standpoint, is the most important part of $L_{II}$: namely that the contribution of each individual force is a change in motion in the direction of the force, proportional to its strength. But FMA entails more besides. Unlike $L_{II}$, it clearly does entail that a body's overall change in motion is a function of the forces impressed on it, so that every change in a body's state of motion is bound to be due to a force. Hence FMA implies the First Law, on both the strong and weak interpretations. Conversely, the strong First Law $S_I$ and the Second Law $L_{II}$ jointly entail FMA — this follows from Newton's Corollaries 1 and 2, discussed above. So FMA in fact combines the empirical content of Newton's First and Second Laws.[12]

This explains why it is that the weak reading of Newton's First Law, while it is a substantive misunderstanding, was relatively inconsequential in another sense. It did not lead to misconceptions about the predictions and empirical content of Newtonian mechanics. This is thanks to the fact that, through the magic of vector calculus, the Eulerian FMA principle picked up the slack, doing the combined work of both the Second Law $L_{II}$ and the strong First Law $S_I$.

It is also why two of the textbooks I mentioned quite rightly omit all mention of a First Law, and make do with FMA on its own. Depending on the details, you may not need a separate Third Law either (Lindsay and Margenau 1957, §3.4). Thus FMA yields a very elegant alternative formulation of Newtonian Mechanics — no wonder this approach has gained currency. After all, there is nothing sacred about Newton's own particular axiomatisation of his theory of motion, or about using three axioms to do it. But the existence of a more economic, vector-based alternative to Newton's formulation does not raise any exegetical puzzle about Newton's own, tripartite approach. We just need to be clear that FMA is not really Newton's Second Law of Motion, but rather a combination of his First and Second Laws.

(Parenthetically, while Newton's formulation is not sacred, I do think Kuhn was right to praise its pedagogical qualities. FMA combines the negative work of the First Law and the positive work of the Second Law into a single statement. That is mathematically neat, but conceptually confusing. The negative message is especially at risk of getting lost. For it is not just Aristotle and Buridan who believe in natural motions, impetus and the like — school children have strikingly similar beliefs, which for the most part persist into adulthood: see diSessa 1982, McCloskey 1983. Teaching the First Law separately, as a converse to the Second, is a great opportunity to address those pervasive misconceptions head on.)

---

[12] FMA does not capture every aspect of $L_I$ and $L_{II}$. In particular, it fails to express the asymmetry of the relationship between force and momentum change. According to $L_{II}$, forces *compel* changes in motion, and not vice versa. To put the point differently, the force *causes* the change in motion, but the change in motion does not cause the force (Pearl 2000, 338). Or differently still, the force *explains* the change in motion, but the change in motion does not explain the force (Rosen 2017, 280-1). The loss of this asymmetry in the Second Law opened the door to the positivist reinterpretation of $\mathbf{F} = m \cdot \mathbf{a}$ as a *definition* of force. But that is a story best left for another day, and I do not want to dwell on this rather metaphysical point here.

## IV. Origins of the Weak Reading

In Section I, we saw that Newton's own presentation of the First Law in the *Principia* appears explicitly designed to rule out a weak reading of the law. If I am right about this, then how did this weak reading gain such wide acceptance? There is more than one reason, I think. One factor is that, as we just saw, the misreading of the First Law did not lead to misconceptions about the empirical content of Newtonian mechanics, thanks to Euler's FMA principle. Another factor is that, as discussed in Section II, the weak reading is still strong enough to capture what is perhaps the central innovation of Newtonian mechanics: to establish uniform rectilinear motion as the default state of motion. Those circumstances enabled and sustained the tradition that kept the weak reading of the First Law dominant over time. But how did the tradition get started in the first place?

While it is difficult to answer such a question with any great confidence, I think there is a plausible explanation in this case. For most of its history, the chief form in which Newton's *Principia* has been consumed, at least in the English-speaking world, is not in the original Latin, but in the English translation of Andrew Motte. Newton himself never saw this translation, which was likely prepared without his knowledge or permission. It appeared two years after his death, in 1729.[13] Motte's translation contains a mistake in its rendering of the First Law (Newton 1729, 1846, my emphasis):

> Every body perseveres in its state of rest, or of uniform motion in a right line, ***unless*** it is compelled to change that state by forces impress'd thereon. (M$_I$)

The crucial departure from the Latin[14] is the omission of the word *quatenus*, which means "insofar as" or "to the extent that." By substituting "unless" for Newton's "except insofar" (*nisi quatenus*), Motte changed the content of the statement. In Newton's formulation, the second clause limits the extent to which bodies that are subject to forces diverge from motion in a straight line. But in Motte's version, the second clause instead appears to exempt bodies subject to forces from the requirement of preserving their state of motion. The revised edition of the Motte translation by Florian Cajori retains the "unless" substitution: "every body continues in its state of rest or of uniform motion in a right line, *unless* it is compelled to change that state by forces impressed upon it" (Newton 1962).

Besides that of Andrew Motte, only two other English translations of the *Principia* have ever been published. One is Robert Thorp's of 1777, which is correct on the crucial point: "Every body perseveres in its state of resting, or of moving uniformly in a right line, *as far as it* is not compelled to change that state by external forces impressed upon it." However, Thorp's edition never achieved the canonical status of the Motte translation: it covered only the first volume of the work, and is mostly noted for its commentary (Cohen 1999, §2.1). The only other published translation is the Cohen and Whitman edition cited above, which has "except insofar". This edition came out in 1999. Neither of these alternative translations can hold a candle to the pervasive impact of Motte's rendition of the text, which is still the most readily available online version of the *Principia* today.

---

[13] I. Bernard Cohen (1963) has a history of the Motte translation. The edition was the initiative of the translator's brother, publisher and bookseller Benjamin Motte. The evidence indicates that the Motte brothers never received Newton's permission for the translation, or indeed had any contact with him at all (Cohen 1963, 348). Cohen claims that "it would not have been out of character for Benjamin Motte to commission a translation from his brother while Newton was still alive, in order to have it ready for publication as soon as possible after" (325).

[14] *Corpus omne perseverare in statu suo quiescendi vel movendi uniformiter in directum, nisi quatenus illud a viribus impressis cogitur statum suum mutare*. See also footnote 2 above.

So it should be no surprise that over the course of three centuries, Motte's "unless" made its way into countless physics textbooks at every level (Galili and Tseitlin 2003), into translations in other languages,[15] as well as into history books, classrooms and lecture halls all over the world, solidifying and canonising the tradition of understanding the First Law as a claim about force-free bodies. Though the "except insofar" translation is not entirely unheard of, Motte-style "unless" formulations of the First Law remain dominant by far, as an internet search for the phrase "Every body continues" will abundantly confirm.

Moreover, it is not widely recognised that the "except insofar" translation not only captures Newton's wording better, but also makes a substantive difference to the content of the law. Consider the extended introduction to the Cohen-Whitman edition of the *Principia*, which contains a detailed catalogue of subtle translation errors in the Motte and Motte-Cajori editions (Cohen 1999, Ch. 2). The "unless" substitution does not even get a mention there. This suggests that Cohen and Whitman themselves considered the difference between Motte's translation and their own rendition of the First Law to be merely cosmetic.

Another case in point is a detailed discussion of the First Law by the great historian of science Alexandre Koyré (1965, 66-9). After quoting the Motte-Cajori translation of the First Law, Koyré warns us that the Latin "expresses Sir Isaac's thought much better than the modern English translation," and that "every word of this formulation is important, both *in se* and for Newton who, as we now know, was an extremely careful writer, who wrote and rewrote the same passage, sometimes five or six times, until it gave him complete satisfaction." He continues with a series of insightful points, opining on the correct translation of *perseverare*, and reflecting on the historic importance of the phrase *in directum*. But what is curiously absent from Koyré's surgical examination of this translation of the First Law is any consideration of the word *quatenus*, "insofar as".[16] In light of the cited remark, it seems ironic that Koyré fails to take note of the fact that this word is entirely omitted from the translation.

Again, this oversight should be understood in light of hundreds of years of tradition — it attests once more to the firm grip that the weak reading has had on our understanding of Newton. Koyré's adherence to that reading is also apparent from his remark that "The motion dealt with in [the first] law is not the motion of the bodies of our experience; we do not encounter it in our daily lives" (1950, 260). This is a reference to the *Triviality Problem*, a difficulty that arises only on a weak reading of the First Law. It will be worth our while to discuss this problem and its reception in more detail. For it offers a window onto the manifold ways in which the weak reading of the First Law has shaped our scholarly, scientific, and philosophical understanding of inertia.

---

[15] The 1759 French translation by the marquise Émilie du Châtelet has "à moins que" ("unless"). The German and Italian translations achieve the same effect with "wenn nicht" ("when not") and "se non" ("if not") respectively (Newton 1759, 1872, 1925).

[16] Koyré does pause to examine the Cartesian phrase "so far as it is able" (*quantum in se est*) (Ibid. 70, 75-6), but does not note the parallel with *nisi quatenus* in the First Law (discussed in Section I). I.B. Cohen was still more emphatic about the importance the phrase *quantum in se est* (Cohen 1964), without apparently conferring comparable significance on its mirror image in the First Law.

## V. The Triviality Problem and the Influence of the Weak Reading

One way to introduce the Triviality Problem is with a famous joke by Arthur Eddington. A skeptical student interrogates their physics teacher about the principle of inertia, defending the view that motion is something that naturally exhausts itself. "After all," says the student, "a bicycle stops of its own accord unless you impress a force to keep it going." The teacher retorts that the bicycle is subject to resisting forces which slow it down, and points out that a stone skimming over ice maintains its uniform rectilinear motion for longer. "Ah," says the student "I see what you mean. So if the stone were projected into a vacuum, without any resistance, then it would continue uniformly in a straight line." "No," replies the teacher, "then it would follow a downward parabola." "But there is nothing to pull it down!" exclaims the student. "There is," explains the teacher, "the invisible hand of gravity compels it to change its state." And so on. Finally, the student becomes exasperated: "So when *does* a body keep going in a straight line?" Replies the teacher: "*Every* body continues in its state of rest or uniform motion in a straight line, except insofar as it doesn't" (Eddington 1929, 123-5).

The point of the joke is that strictly speaking, there are no force-free bodies. So strictly speaking there is no inertial motion either. This is perfectly true, by Newton's own lights: if only because of universal gravitation, every massive body is acted on by *some* force or other, provided there is at least one other massive body in the universe. So the weak reading $W_I$ of the First Law has no real-world application at all. As a generalisation with no instances, $W_I$ is technically true, in the logician's sense. But in that technical sense, it would be no less true to say that all force-free bodies move in spirals while singing the Russian national anthem. Surely there must be more to the First Law than truth in this trivial sense. From the perspective of the weak reading, the question then becomes: what more is there? What is it that the First Law is *really* telling us? This is the Triviality Problem.

Note that Eddington's joke takes place in a high school classroom. This is no accident. It lets us in on the fact that the Triviality Problem is not an abstruse issue for academics, but something that critically inclined high school students quite naturally come to grapple with. In this way, the weak reading of the First Law has lent urgency and mystery to foundational and conceptual questions around inertia even in early physics education. This might go some way towards explaining why the puzzle had such a tremendous impact, giving rise to such a wide variety of responses.

Probably the most influential response is to recast the First Law as a disguised operational definition of *inertial frames* (Lange 1885/2014, Torretti 1983). This idea made its way into many physics textbooks (e.g. Landau et al. 1969, Reif 1995; see also Galili and Tseitlin 2003, 64). Relatedly, Julian Barbour views the First Law as an operational definition of *time duration* — a "cosmic clock" (Kelvin and Tait 1883; Barbour 2001, Ch. 6). Ellis 1965 and Arons 1990 argue that the First Law is a partial operationalisation of the concept of *force*. The idea there is that, in the Newtonian paradigm, impressed forces are posited whenever a body's motion is *not* uniform and rectilinear. Treating the First Law as a definition, these responses all embrace the triviality of the First Law, emphasising its conceptual and heuristic aspects over its empirical content.

Besides this operational approach to the Triviality Problem, there are the geometric approach, the modal approach and the historical approach. The geometric approach rests on the observation that the weak First Law implies the existence of infinite, straight spacetime trajectories through any point and in any direction (Earman and Friedman 1973, Anderson 1990, Brown 2006). By treating this as a *consequence* rather than a *presupposition* of the First Law, the law is transformed into a load-bearing principle in the theory of spacetime structure. The modal approach argues that even if the First Law has no real-world application, it still applies to counterfactual or imaginary situations where forces are absent (Koyré 1950, 260-1; Cohen 1980, 106-9; Feynman et al. 1963, §6.2). Hay 1957 and Hanson 1965 spell this out in terms of counterfactual conditionals: the First Law tells us what *would* happen if no forces *were* acting. Closely related is Pfister's (2004) view that the First Law is an idealisation that real situations only approximate. Finally, the historical approach locates the First Law's significance in its discontinuity with prior theories of mechanics (Maudlin 2001, Koyré 1965, Cohen 2002).

There is one other response to the Triviality Problem I wish to mention, namely that of Henri Poincaré. This is the closest antecedent I could find to my own solution. After introducing the problem, Poincaré grants that the weak First Law or principle of inertia is not testable. He then proposes that it should instead be viewed as an instance of a stronger principle that does have verifiable consequences (note again the classroom setting):

> Teachers of mechanics usually … add that the principle of inertia is verified by its consequences. This is very badly expressed; they evidently mean that various consequences may be verified of a more general principle, of which the principle of inertia is only a particular case. I shall propose for this general principle the following enunciation:— The acceleration of a body depends only on its position and that of neighbouring bodies, and their velocities. (Poincaré 1905, 92)

The general principle Poincaré articulates here is a close cousin to the strong First Law $S_I$, which says that acceleration depends only on the forces impressed on a body. If we assume that those forces in turn depend "only on its position and that of neighbouring bodies, and their velocities," this yields Poincaré's generalisation. Poincaré does not suggest, however, that the principle of inertia really is this stronger, more general principle.

If the First Law really is the strong principle $S_I$, then there simply is no Triviality Problem. After all, the strong First Law $S_I$ says that *all* bodies diverge from their uniform rectilinear motions only insofar as forces compel them to. There are lots of real-world instances of this principle, including the examples Newton gives of a projectile, a spinning top, and planetary motion. In fact, the trajectory of any Newtonian body is an instance of $S_I$. Or looking at it a different way, the strong First Law says that every *change* in a body's state of motion is due to a force — and that generalisation has as many instances as there are changes in bodies' states of motion. Far from being a trivial truth, the strong First Law is arguably *false* in light of General Relativity: in curved spacetimes, particles change direction without the action of any impressed force.

We are forced to conclude that in some sense, the literature covered in the above survey is full of solutions to a non-existent problem. Surprisingly though, this barely detracts from the interest of these bodies of work. Most of the studies cited do not have the Triviality Problem as their sole motivation, and were always intended as exercises in rational reconstruction more than Newton exegesis. They raise and answer important questions about the modality of natural law, about idealisation, about the role of operational definitions and geometry in physics, and about the historical significance of the principle of inertia. I think the importance and significance of those questions survives the dissolution of the Triviality Problem.

This makes the Triviality Problem something of a happy accident. From an exegetical point of view, the problem is based on a misunderstanding. But one would be hard-pressed to deny that it nonetheless proved to be an immensely fruitful guide to further inquiry, shining a spotlight on a series of fascinating questions that are independently worthwhile.

On the received view, Newton's First Law of Motion is the principle that bodies move uniformly in straight lines, as long as there are no forces acting on them. For a long time, this traditional reading of the law enjoyed a state of fairly uniform adherence. Three forces compel us to change that state. The first is Newton's wording of the law itself, and especially his use of the phrase "except insofar". The second is Newton's elucidation of his First Law, which includes applications to bodies whose trajectories are manifestly not rectilinear, like the parts of a spinning top. The third is the fact that the traditional reading is not strong enough to yield a complete formulation of Newtonian Mechanics. Acting conjointly, these three forces carry us towards a different view. On this view, the First Law is a stronger, more general principle, governing the extent to which a body, *any* body, *diverges* from uniform rectilinear motion. The First Law limits how much a body's speed and direction can change, stating that these only change to the extent that the forces impressed on the body compel them to change. In other words, a body's state of uniform, rectilinear motion never changes of its own accord: it's forced changes only.